\documentclass[prx,aps,twocolumn,longbibliography, superscriptaddress]{revtex4-2}

\pdfoutput=1

\newcommand{\tb}{t_{\textrm b}}
\newcommand{\wb}{\omega_{\textrm b}}
\newcommand{\Eb}{E_{\textrm b}}
\newcommand{\kB}{k_{\textrm{B}}}

\newcommand{\UD}{U_{\textrm{td}}}

\newcommand{\Nobs}{N_{\textrm{obs}}}
\newcommand{\Bres}{B_{\textrm{res}}}
\newcommand{\abg}{a_{\textrm{bg}}}
\newcommand{\avdw}{r_{\rm vdW}}

\usepackage{float}

\usepackage{fourier}
\usepackage{amsmath,amssymb}
\usepackage{mathtools}

\usepackage{times}
\usepackage{upgreek}
\usepackage{psfrag} 
\usepackage{latexsym} 
\usepackage{amstext}
\usepackage{amsxtra} 

\usepackage{dcolumn}

\usepackage{textcomp}
\usepackage{amsfonts}
\usepackage{graphicx}
\usepackage{bm}
\usepackage{color}
\usepackage{braket}
\usepackage{units}
\usepackage[svgnames]{xcolor}

\definecolor{myColor}{rgb}{0.02,0.12,0.3}
\definecolor{myciteColor}{rgb}{0.39,0.7,0.89}
\usepackage[colorlinks=true,citecolor=myColor,linkcolor=myColor,urlcolor=myColor]{hyperref}

\def\be{\begin{equation}}
\def\ee{\end{equation}}

\usepackage{amsthm, amssymb}
\usepackage{empheq}
\usepackage{cases}

\def\nobreakbefore{%
  \relax\ifvmode\else
    \ifhmode
      \ifdim\lastskip > 0pt\relax
        \unskip\nobreakspace
      \else 
        \nobreakspace
      \fi
    \fi
  \fi
}
\let\oldcite\cite
\renewcommand\cite{\nobreakbefore\oldcite}

\begin{document} 
\title{
Pinpointing Feshbach Resonances and Testing Efimov Universalities in $^{39}$K\\
}
\author{Ji\v{r}\'{i}~Etrych}
\affiliation{Cavendish Laboratory, University of Cambridge, J. J. Thomson Avenue, Cambridge CB3 0HE, United Kingdom}
\author{Gevorg~Martirosyan}
\affiliation{Cavendish Laboratory, University of Cambridge, J. J. Thomson Avenue, Cambridge CB3 0HE, United Kingdom}
\author{Alec~Cao}
\affiliation{Cavendish Laboratory, University of Cambridge, J. J. Thomson Avenue, Cambridge CB3 0HE, United Kingdom}
\author{Jake~A.~P.~Glidden}
\affiliation{Cavendish Laboratory, University of Cambridge, J. J. Thomson Avenue, Cambridge CB3 0HE, United Kingdom}
\author{Lena~H.~Dogra}
\affiliation{Cavendish Laboratory, University of Cambridge, J. J. Thomson Avenue, Cambridge CB3 0HE, United Kingdom}
\author{Jeremy~M.~Hutson}
\affiliation{Joint Quantum Centre (JQC) Durham-Newcastle, Department of Chemistry, Durham University, South Road, Durham DH1 3LE, United Kingdom}
\author{Zoran~Hadzibabic}
\affiliation{Cavendish Laboratory, University of Cambridge, J. J. Thomson Avenue, Cambridge CB3 0HE, United Kingdom}
\author{Christoph~Eigen}
\email{ce330@cam.ac.uk}
\affiliation{Cavendish Laboratory, University of Cambridge, J. J. Thomson Avenue, Cambridge CB3 0HE, United Kingdom}

\begin{abstract}
Using a combination of bound-state spectroscopy and loss spectroscopy, we pinpoint eight intrastate Feshbach resonances in $^{39}$K, as well as six previously unexplored interstate ones. We also perform a detailed characterization of four of the intrastate resonances and two of the interstate ones.
We carry out coupled-channel scattering calculations and find good agreement with experiment.
The combination of experiment and theory provides 
a faithful map of the scattering length $a$ and permits precision measurements of the signatures of Efimov physics across four intermediate-strength resonances.
We measure the modulation of the $a^4$ scaling of the three-body loss coefficient for both $a<0$ and $a>0$, as well as the many-body loss dynamics at unitarity (where $a$ diverges). The absolute positions of the observed Efimov features confirm a ubiquitous breakdown of Efimov--van-der-Waals universality in~$^{39}$K, while their relative positions are in agreement with the universal Efimov ratios.
The loss dynamics at the three broadest Feshbach resonances are universal within experimental uncertainties, consistent with observing little variation in the Efimov width parameters.

\end{abstract}
\maketitle 
\section{Introduction}

The exquisite ability to tune interatomic interactions using magnetic Feshbach resonances lies at the heart of many ultracold-atom experiments~\cite{Chin:2010}. A single resonance can provide access to the full range of contact interactions, characterized by the s-wave scattering length $a$.
Near resonance, strongly correlated phases of matter~\cite{Bloch:2008} can be realized, with highlights including
unitary Bose ~\cite{Rem:2013,Fletcher:2013,Makotyn:2014,Eismann:2016,Fletcher:2017,Klauss:2017,Eigen:2017,Eigen:2018} and Fermi\cite{Inguscio:2007,Zwerger:2011,Zwierlein:2014}
gases.
Focusing on Bose gases, the weakly repulsive regime offers a textbook setting for exploring interacting Bose--Einstein condensates\cite{Pethick:2002,Pitaevskii:2016},
while tuneable attractive interactions facilitate studies of condensate collapse\cite{Roberts:2001,Donley:2001,Altin:2011, Eigen:2016}, soliton formation~\cite{Strecker:2002,Khaykovich:2002,Cornish:2006,Nguyen:2017,Chen:2020}, and negative absolute temperatures\cite{Braun:2013}.
The zero crossing of $a$ provides pristine conditions for simulating noninteracting quantum phenomena~(see \emph{e.g.}~\cite{Roati:2008,Fattori:2008a}). 
Knowledge of the details of Feshbach resonances and the resulting map of scattering lengths across multiple atomic states facilitates studies of quantum mixtures\footnote{Similarly, navigating the zoo of Feshbach resonances\cite{Frisch:2014} in magnetic lanthanides has been crucial for controlling the interplay between dipolar and contact interactions\cite{Norcia:2021}.} and many-body interferometry~\cite{Cetina:2016,Fletcher:2017,Zou:2021}.

Feshbach resonances also provide a testbed for fundamental few-body physics, with the resonance positions $\Bres$ acting as invaluable benchmarks.
Sufficiently close to resonance, the structure of few-body bound states is particularly simple: a dimer of size $a$ exists on the repulsive side ($a>0$) and becomes unbound as the magnetic field $B\to \Bres$. This provides a gateway into the realm of ultracold molecules~\cite{Koehler:2006,Carr:2009,Bohn:2017}, and a particularly accurate method for pinpointing $\Bres$\cite{Chin:2010,Zurn:2013,Sala:2013,Chapurin:2019}.
In Bose gases, the Efimov effect\cite{Efimov:1970,Braaten:2007,Greene:2017,Naidon:2017,DIncao:2018} leads to a spectrum of three-body bound states, persisting even when the dimer is unbound.

A hallmark of Efimov states is their discrete scaling symmetry, which manifests in the log-periodic modulation of few-body observables, such as the three-body recombination rate \cite{Kraemer:2006,Knoop:2009,Pollack:2009,Zaccanti:2009,Gross:2009,Berninger:2011,Wild:2012,Xie:2020}
(see also~\footnote{Efimov universalities have also been experimentally explored in systems with unequal atomic mass or distinguishable spin states~\cite{Barontini:2009,Williams:2009,Lompe:2010,Helfrich:2011,Bloom:2013,Pires:2014,Maier:2015,Johansen:2017}.}).
While the functional form of these modulations is predicted by Efimov theory, their phase (captured by $a_-$, where the lowest Efimov state meets the continuum) depends on the details of the short-range interaction.
Remarkably, for many Feshbach resonances, across spin states and atomic species, $a_-$ was measured to be within $20\%$ of $-9\,\avdw$~\cite{Berninger:2011,Johansen:2017,Mestrom:2017}, where $\avdw$ is the van-der-Waals  length~\cite{Chin:2010}. This Efimov--van-der-Waals universality was traced to the universal form of the interaction potential, predicting $a_-\approx-9.7\avdw$ for broad resonances\cite{Wang:2012,Naidon:2014}.

The Efimov effect is posited to influence the many-body state at unitarity as well, from setting the lifetime of the unitary gas~\cite{Rem:2013,Eismann:2016}
to predictions of a low-temperature superfluid of Efimov trimers~\cite{Piatecki:2014, Musolino:2022}.
So far, three-body correlations in a thermal unitary Bose gas have been observed~\cite{Fletcher:2017}, but in the degenerate case signatures are limited to experiments that convert the strongly correlated state into an atom-molecule mixture by sweeping from unitary to weak interactions\cite{Klauss:2017,Eigen:2017}.

Over the last 15 years $^{39}$K has proven to be a versatile atom for quantum-gas experiments.
Numerous $^{39}$K Feshbach resonances were surveyed early on~\cite{Roati:2007,DErrico:2007,Roy:2013}, with recent precision measurements of a select few~\cite{Fletcher:2017,Tanzi:2018,Chapurin:2019}.
The intermediate-strength nature of its broad resonances has rendered $^{39}$K ideal for tests of the Efimov universalities\cite{Zaccanti:2009,Roy:2013,Chapurin:2019,Xie:2020}.
Recently, a high-precision measurement on a $33.6$\,G resonance revealed a breakdown of Efimov--van-der-Waals universality\cite{Chapurin:2019}, finding $a_-=-14.05(17)\avdw$, while the Efimov ratios were largely consistent with universal theory~\cite{Xie:2020}.

\begin{figure*}[t!]
\centerline{\includegraphics[width=\textwidth]{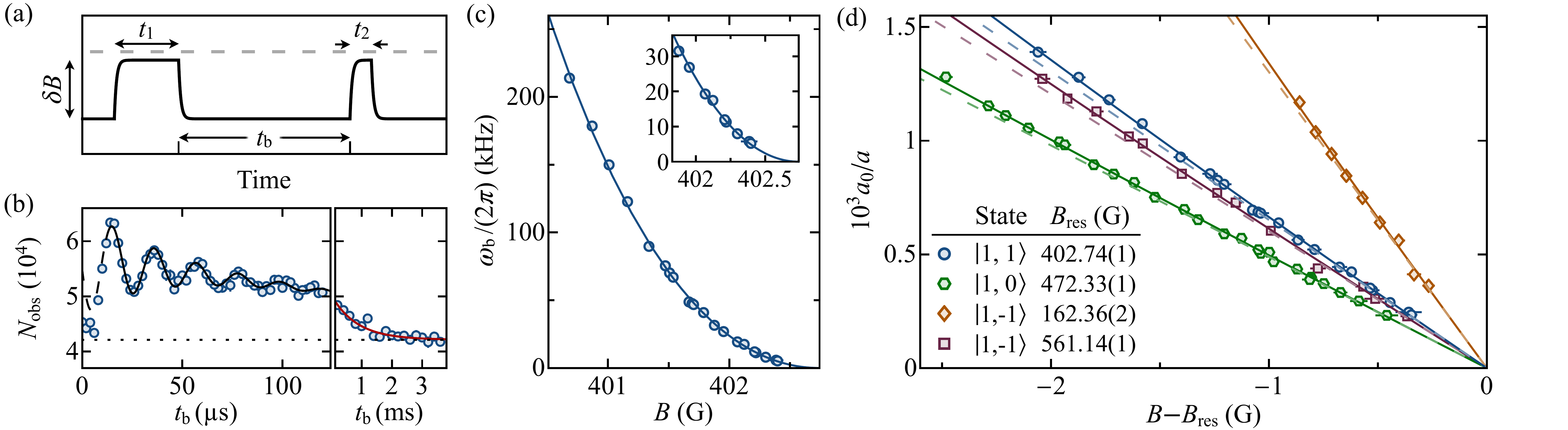}}
\caption{
High-precision bound-state spectroscopy in $^{39}$K. (a)~Quench protocol used to probe the Feshbach dimer binding energy $\Eb$  interferometrically.
We perform two field pulses of magnitude $\delta B$ towards $B_{\rm res}$ (dashed line), separated by a hold time $\tb$ at the magnetic field of interest $B$.
(b) Example of the evolution of the observed atom number $\Nobs (\tb)$. We extract the oscillation frequency $\wb$ by fitting  the early-time data (black line, see text for details). At long times, $\Nobs$ slowly decays due to molecular loss processes; the red line shows a guide to the eye and the dotted line the long-time asymptote.
(c) $B$-field dependence of $\wb$ for the $|1,1\rangle$ state Feshbach resonance near $402.7$\,G. The inset shows a zoom-in near resonance, and we use Eqs.~(\ref{eqaB},\ref{eqEb}) to extract $\Bres$ (see text for details).
(d) Characterization of four intrastate resonances (see legend and Table~\ref{tab1}). Inverse scattering length $1/a$, calculated from the measured $\wb$, versus detuning from resonance $B-\Bres$. The solid lines show fits using Eqs.~(\ref{eqaB},\ref{eqEb}), while the dashed lines show the linear approximation valid close to resonance.}
\vspace{-0.2em}
\label{fig1}
\end{figure*}

In this article, we use a combination of bound-state and loss spectroscopy to pinpoint and characterize eight intrastate Feshbach resonances in $^{39}$K and six previously unexplored interstate ones.
We compare the properties of the resonances with the results of coupled-channel scattering calculations.
Using the resultant accurate determination of the interaction landscape and leveraging the advantages of homogeneous samples~\cite{Gaunt:2013,Navon:2021}, we map out Efimov loss extrema across four broad Feshbach resonances. 
Our measurements reveal a ubiquitous breakdown of Efimov--van-der-Waals-universality~\cite{Mestrom:2017,Johansen:2017}, consistently finding $a_-$ in the range $-13(1)\avdw$, which we attribute to the similar intermediate-strength resonance characters.
The Efimov ratios between features across the Feshbach resonance are in agreement with Efimov theory. 
We also explore the many-body loss dynamics in unitary gases for the three broadest Feshbach resonances in different spin states. We observe universal behavior, consistent with measuring similar Efimov width parameters $\eta^*$.
\vspace{-1em}
\section{Bound-state Spectroscopy}
\vspace{-0.5em}
In the vicinity of an isolated Feshbach resonance, neglecting inelasticity, $a$ approximately follows \cite{Moerdijk:1995b}
\begin{equation}
a=\abg\left( 1-\frac{\Delta}{B-\Bres} \right),
\label{eqaB}
\end{equation}
where $a_\textrm{bg}$ is a slowly varying background scattering length and $\Delta$ is the resonance width. Close to the pole, $a\approx -a_{\rm bg}\Delta/(B-B_{\rm res})$. If $a_\textrm{bg}$ is constant across the entire resonance, $a$ crosses zero at $B_\textrm{res}+\Delta$. More generally, however, the position $B_\textrm{zero}$ of the zero crossing may differ from this value.
In the presence of inelasticity, Eq.\ (\ref{eqaB}) breaks down close to resonance \cite{Hutson:2007}.
However, for the resonances considered here, this breakdown is insignificant  more than 1~mG from $B_\textrm{res}$, as shown theoretically in \hyperref[appA]{Appendix\,A}.

For large positive values of $a$ close to resonance, the binding energy of the least-bound molecular state is\cite{Gribakin:1993}
\begin{equation}
E_{\rm b}=\frac{\hbar^2}{m (a-\bar{a})^2}\,,
\label{eqEb}
\end{equation}
where $m$ is the atom mass, $\hbar$ the reduced Planck's constant, $\bar{a}=[4\pi/\Gamma(1/4)^2]\avdw\approx 61.8a_0$, $\avdw=64.6 a_0$~\cite{Falke:2008} for $^{39}$K, and $a_0$ is the Bohr radius.

For our bound-state spectroscopy we begin with a quasi-pure $^{39}$K Bose--Einstein condensate confined in the uniform potential of a cylindrical optical box trap~\cite{Gaunt:2013,Eigen:2016,Navon:2021} of volume $V\approx3\times 10^{-14}\,\text{m}^3$.
We prepare weakly interacting repulsive spin-polarized samples in a hyperfine state of choice within the \mbox{$F=1$} manifold.
Following~ Refs.~\cite{Donley:2002,Claussen:2003}, and as depicted in \hyperref[fig1]{Fig.\,1(a)}, we employ a Ramsey-like quench protocol. After tuning $B$ to a value of interest over tens of ms~\footnote{Note this ramp is not necessarily adiabatic, which can excite the condensate\cite{Matthews:1998}. We have checked that our $\wb$ measurements are robust to the details of the ramp.
},
we perform two field pulses of magnitude $\delta B$ towards $B_{\rm res}$, separated by a time $\tb$~\footnote{We independently measure $B$ during $t_{\rm{b}}$ using rf spectroscopy performed under identical experimental conditions. Our remaining systematic error on $B$ on these timescales is $\sim10$\,mG.}. We use $t_1$ on the order of tens of $\upmu$s, short $t_2=10\,\upmu$s, and our field changes occur in~$\sim 2\,\upmu$s~\cite{Eigen:2017}.
The first pulse initiates a coherent superposition of atoms and molecules~\footnote{
While such quenches can lead to the formation of both dimers and trimers\cite{Klauss:2017, Eigen:2017}, we find that our measurements are consistent with assuming just a mixture of atoms and dimers.
}. During the evolution time~$\tb$ a phase difference (set by~$E_{\rm b}$) accumulates, before the second pulse projects this difference onto a population imbalance. Finally, we record a time-of-flight absorption image of the sample.

As shown in \hyperref[fig1]{Fig.\,1(b)}, the observed atom number $\Nobs$ exhibits decaying oscillations with $\tb$. Our imaging detects only free atoms, and we associate the initial increase of $\Nobs$ with the number of atoms transferred to the molecular state.
Once the oscillation has decayed, we are still able to revert a significant fraction of molecules to atoms using the second pulse, consistent with a long-lived molecular sample that eventually decays due to molecular loss processes (see \emph{e.g.} \cite{Xie:2020}).

To extract the oscillation frequency $\wb$, which directly gives $E_{\rm b}=\hbar \wb$, we fit $\Nobs(\tb)$ with $N_0 -\alpha \tb + N_{\rm m} \exp(-\tb/\tau) \sin(\wb \tb+\phi)$~\footnote{We typically exclude the first $10\,\upmu$s from the fit to mitigate effects of our finite quench times.
We directly relate the oscillation frequency to $\wb$, neglecting model-dependent frequency shifts due to damping, which we assess to be negligible.}. While most extracted fit parameters are sensitive to the details of the quench protocol, crucially, $\wb$ is robust\cite{Donley:2002,Claussen:2003}.

In \hyperref[fig1]{Fig.\,1(c)} we show $\omega_{\rm b}/(2\pi)$ versus $B$ for the $|1,1\rangle$ state near the $402.7$\,G Feshbach resonance.
The solid line shows a fit to the data using Eqs.\,(\ref{eqaB},\ref{eqEb}), which yields $B_{\rm res}=402.76(3)\,$G, $a_{\rm bg} \Delta = 1600(70)\,a_0$G, and a relatively uncertain $a_{\rm bg}=-60(30) a_0$. If we assume that $\Delta=B_\textrm{zero}-B_\textrm{res}$ and use the independently measured $B_{\rm zero}=350.4(1)$\,G \cite{Fattori:2008b,Eigen:2016}, to constrain the fit, we obtain a refined $B_{\rm res}=402.74(1)\,$G and $a_{\rm bg}=-29.3(3)\,a_0$.

In \hyperref[fig1]{Fig.\,1(d)} we plot $1/a$ versus $B-B_{\rm res}$, and also show analogous measurements across three other resonances. We depict the constrained fits (solid lines) and their linear approximation valid near $\Bres$ (dashed lines).

\begin{figure}[t!]
\centerline{\includegraphics[width=\columnwidth]{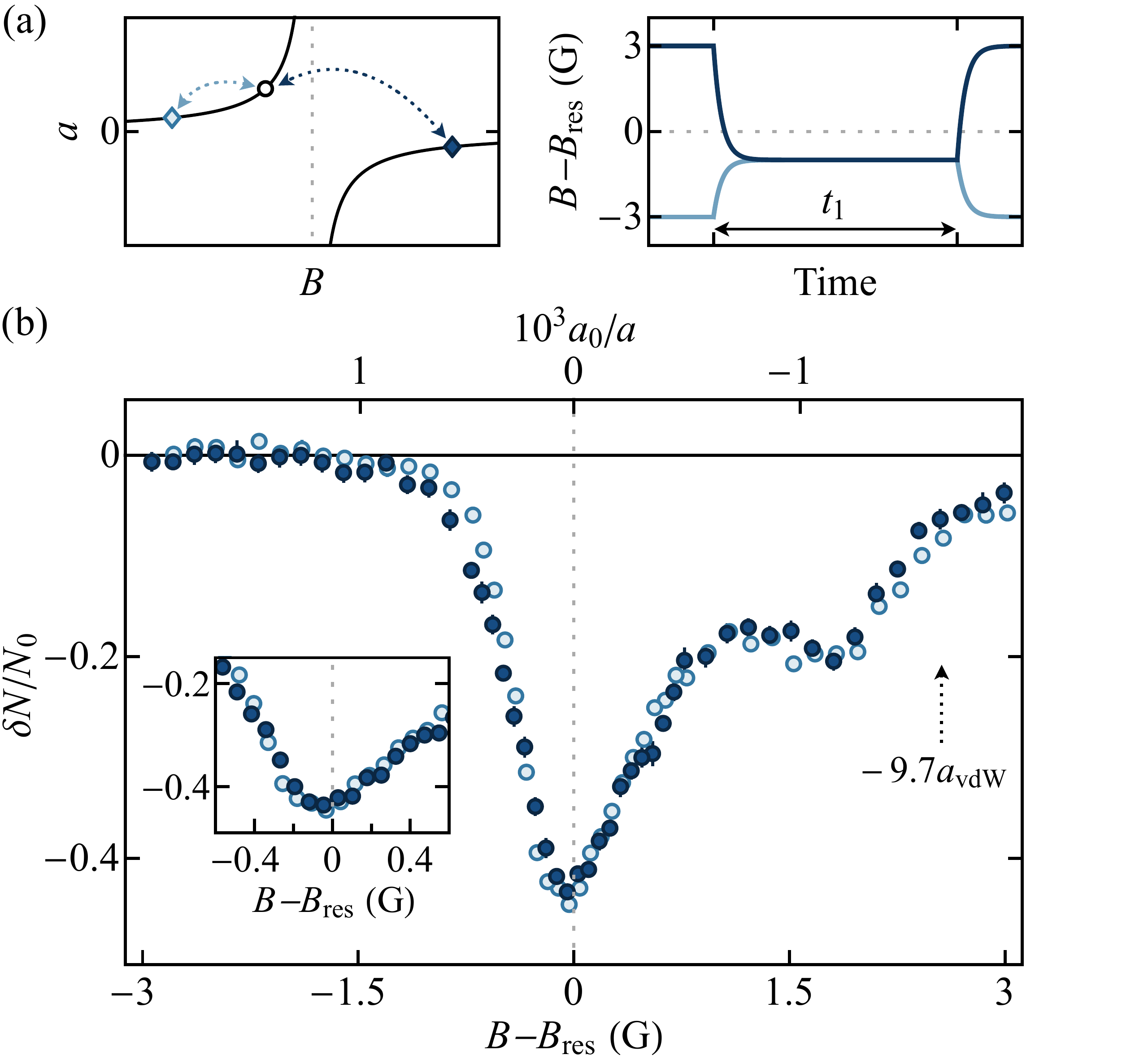}}
\caption{Benchmarking quench-based loss spectroscopy. (a) Interaction landscape and quench protocol. We prepare thermal samples in the $|1,1\rangle$ state $\pm3\,$G away from $\Bres=402.74(1)\,$G (dashed line). In both cases we quench to the same final $B$, wait for a time $t_1$, before quenching back and probing the sample.
(b)~Fractional change in atom number $\delta N/N_0=\Nobs/N-1$ across the Feshbach resonance for $t_1=400\,\upmu$s. The asymmetric primary loss feature (see inset for a zoom-in) occurs at $\Bres$, validating our loss spectroscopy, while the secondary feature arises due to Efimov physics.}
\label{fig2}
\end{figure}

\section{Benchmarking Loss spectroscopy}
\vspace{-0.5em}
We next benchmark the more economical loss spectroscopy as an accurate albeit less precise method for measuring $\Bres$.
We focus on the $|1,1\rangle$ state near $\Bres=402.74(1)$\,G and prepare samples of density $n_0\approx1.3\,\upmu\text{m}^{-3}$ at a temperature $T_0\approx 100\,$nK; thermal samples are used to facilitate starting both above and below resonance. As depicted in \hyperref[fig2]{Fig.\,2(a)}, in both cases we then quench to the same final $B$ and wait for a time $t_1$, before quenching back to measure $\Nobs$.

In \hyperref[fig2]{Fig.\,2(b)} we plot the fractional change in atom number $\delta N/N_0=\Nobs/N_0-1$ after $t_1=400\,\upmu$s as a function of $B-B_{\rm res}$ for both quenches.
We observe maximal loss near $B_{\rm res}$, and see that the quench direction does not matter (this relies on choosing a sufficiently long $t_1$, see 
\hyperref[appB]{Appendix~B}).
The central loss feature is asymmetric, with enhanced loss for $a<0$ and a pronounced secondary peak, as expected from Efimov physics. We associate the secondary peak with the lowest-lying Efimov resonance at $a_-$, which occurs near but visibly closer to $\Bres$ than $a_-=-9.7\avdw$ predicted from Efimov--van-der-Waals universality.
Our measurements establish quench-based loss-spectroscopy as an accurate tool to pinpoint Feshbach resonances up to the width of the loss feature~\footnote{We report the half-width half-maximum (extracted using Gaussian fits) as the error bar. This achieves a precision of $\Bres$ up to a few percent of $\Delta$.}.

\section{Summary of $^{39}$K Feshbach resonances}

In \hyperref[tab1]{Table\,I} we summarize our intrastate resonance measurements, including those from \hyperref[fig1]{Fig.\,1(c)} and four based on loss spectroscopy.
We also include the high-precision measurement from Ref.~\cite{Chapurin:2019} and the high-field one from Ref.~\cite{DErrico:2007} (see also \footnote{Note that several $d$-wave resonances have recently also been located~\cite{Fouche:2019,Tiemann:2020}.}).
Moreover, we include our independent measurements of the associated zero-crossings
\footnote{Note that these measurements are susceptible to several systematic shifts that could arise, including from the nonzero magnetic dipole-dipole interactions~\cite{Fattori:2008b} or due to elastic three-body interactions~\cite{Mestrom:2020}. For the resonance at $162.4$\,G we have conservatively assumed $\Delta=-35(15)$\,G to constrain the fit; a zero-crossing is prevented here due to the resonance at $33.6$\,G. Crucially, our determination of $\Bres$ and $a_{\rm bg}\Delta$ is insensitive to changes in $B_{\rm zero}$.}
based on either the critical condition for collapse~\cite{Roberts:2001,Eigen:2016} or thermalization rates \cite{Roberts:1998}.
Note that our measured values of $a_\textrm{bg}\Delta$ are those that characterize the strength of the pole in $a$, and are not necessarily consistent with independently measured values of $B_\textrm{zero}$ and $a_\textrm{bg}$ far from the pole.

\begin{table}[t!]
\vspace{-1em}
\caption{\label{tab1}
Summary of intrastate Feshbach resonance measurements in $^{39}$K: the resonance position $\Bres$, effective width $\abg \Delta$, and zero crossing $B_{\rm zero}$. We also include the measurements from Refs.~\cite{DErrico:2007,Fattori:2008b,Chapurin:2019}, as well as the corresponding atomic-state magnetic moments $\bar{\mu}$ (at $\Bres$), which are relevant for experiments. The `-' denote cases where the parameter has not been experimentally determined, while `/' indicate those that do not occur.}
\begin{ruledtabular}
\begin{tabular}{ccccc}
$\ket{F,m_{F}}$ & $B_{\rm res}$ (G) & $a_{\rm bg} \Delta$ ($a_0$\,G) & $B_{\rm zero}$ (G) & $\bar{\mu}  (\mu_{ B})$ \\
\hline
$\ket{1,1}$&$25.91(6)$ & - & - & $-0.605$ \\
$\ket{1,1}$&$402.74(1)$ & $1530(20)$ & $350.4(1)$\cite{Fattori:2008b} & $-0.961$ \\
$\ket{1,1}$&$752.3(1)$\cite{DErrico:2007} & - & - & $-0.987$  \\
$\ket{1,0}$ & 58.97(12)  & - & - & $-0.337$  \\
$\ket{1,0}$ & 65.57(23) & - & - & $-0.370$  \\
$\ket{1,0}$ & $472.33(1)$  & $2040(20)$ & $393.2(2)$ & $-0.945$\\
$\ket{1,0}$ &$491.17(7)$ & $140(30)$\footnotemark[1] & $490.1(2)$ & $-0.949$\\
$\ket{1,-1}$& $33.5820(14)$\cite{Chapurin:2019}   & $-1073$\cite{Chapurin:2019} & / & $\hphantom{-}0.324$ \\  
$\ket{1,-1}$&$162.36(2)$ & $760(20)$ & / & $-0.489$\\
$\ket{1,-1}$& $561.14(2)$ & $1660(20)$ & $504.9(2)$ & $-0.959$  \\
\end{tabular}
\end{ruledtabular}
\footnotetext[1]{See \hyperref[appC]{Appendix\,C}.}
\end{table}

\begin{table}[t!]
\vspace{-1em}
\caption{\label{tab2}
Summary of interstate Feshbach resonance measurements in $^{39}$K, including one from Ref.~\cite{Tanzi:2018}.
We also include the corresponding atomic-state magnetic moments $\bar{\mu}_{1,2}$ (at $\Bres$), relevant for experiments. The `-' denote cases where the parameter has not been experimentally determined.
}
\begin{ruledtabular}
\begin{tabular}{ccccc}
 $\ket{F,m_{F}}_1+ \ket{F,m_{F}}_2$\hphantom{$-$}  & $B_{\rm res}$ (G) & $\abg \Delta$ ($a_0$\,G) & $\bar{\mu}_1 (\mu_{ B})$ & $\bar{\mu}_2 (\mu_{ B})$ \\
\hline
$\ket{1,1}+\ket{1,0}$\hphantom{$-$} &$25.81(6)$ & - & $-0.605$ & $-0.155$ \\
$\ket{1,1}+\ket{1,0}$\hphantom{$-$}  &$39.81(6)$ & -  & $-0.651$ & $-0.235$  \\
$\ket{1,1}+\ket{1,0}$\hphantom{$-$} &$445.42(3)$\footnotemark[1] & $1110(40)$\footnotemark[1] & $-0.967$ & $-0.939$ \\
$\ket{1,1}+\ket{1,-1}$ &$77.6(4)$ & -  & $-0.747$ & $\hphantom{-}0.034$  \\
$\ket{1,1}+\ket{1,-1}$ &$501.6(3)$ & -  & $-0.973$ & $-0.948$  \\
$\ket{1,0}+\ket{1,-1}$ &$113.76(1)$~\cite{Tanzi:2018} & $715(7)$~\cite{Tanzi:2018}  & $-0.569$ & $-0.215$\\
$\ket{1,0}+\ket{1,-1}$ &$526.21(5)$\footnotemark[1] & $970(50)$\footnotemark[1]  & $-0.956$ & $-0.953$  \\
\end{tabular}
\end{ruledtabular}
\footnotetext[1]{See \hyperref[appD]{Appendix\,D}.}
\end{table}

\begin{figure*}[t!]
\centerline{\includegraphics[width=\textwidth]{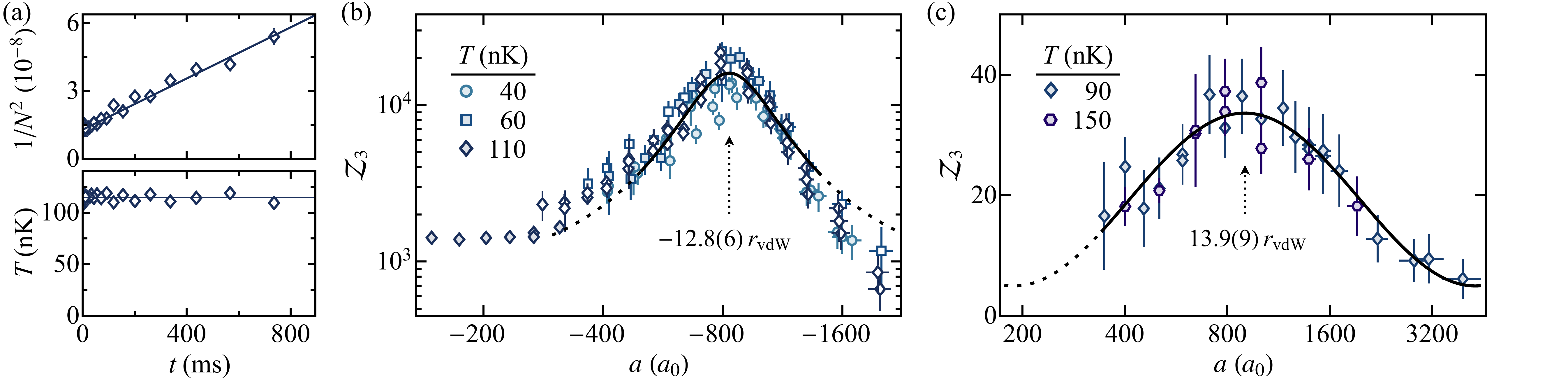}}
\caption{
Testing Efimov universalities in the $|1,1\rangle$ state ($B_{\rm res}=402.74$\,G).
(a) Loss dynamics in a box-trapped thermal gas at $a=-930(30)a_0$, plotting the evolution of $1/N^2$ (top) and temperature $T$ (bottom).
As expected for three-body recombination in a uniform system, $1/N^2(t)$ is linear and $T(t)$ constant.
(b,c) Modulation of the three-body loss coefficient~$\mathcal{Z}_3=m L_3/(3 \hbar a^4)$ for $a<0$~(b) and $a>0$~(c). Different data series are grouped by their approximate $T$ (see legend) and the symbol filling denotes the preparation protocol: $B$-field ramps (closed) and spin flips (open).
Our error bars combine fit errors and the uncertainty in $a$, but exclude a systematic density uncertainty $\lesssim 30\%$.
We extract the positions of the Efimov features $a_-$~(b) and $a_p$~(c) with fits using Eqs.~(\ref{eqZ3}) and a free amplitude (solid lines).
While the ratio $a_p/a_-=-1.08(9)$ is consistent with Efimov universality ($a_p/a_-=-1)$, the absolute positions exclude $a_-=-9.7\avdw$ predicted from Efimov--van-der-Waals universality.
}
\label{fig3}
\end{figure*}

In \hyperref[tab2]{Table\,II} we summarize our interstate resonance measurements, also including one from Ref.~\cite{Tanzi:2018}.
By transposing loss spectroscopy to spin mixtures we locate six previously predicted~\cite{Lysebo:2010} but experimentally elusive interstate resonances. The resonances at $445.4$\,G and $526.2$\,G are promising candidates for studies of quantum mixtures~(\emph{e.g.}~Bose polarons~\cite{Jorgensen:2016,Hu:2016,Yan:2020}). In particular, the magnetic moments of the two states involved  $(\bar{\mu}_1,\bar{\mu}_2)$ are very similar (differing by only $3\%$ and $0.3\%$, respectively), which opens the door to forming magnetically levitated uniform Bose mixtures with tuneable interparticle interactions. 
We further characterize these two resonances using rf molecular association spectroscopy~\cite{Chin:2010,Zirbel:2008,Weber:2008}; see \hyperref[appD]{Appendix\,D} for details.

We also compare our measurements to coupled-channel scattering calculations using state-of-the-art interaction potentials\cite{Tiemann:2020}; see \hyperref[appA]{Appendix\,A} for details. The calculated resonance positions differ from the experiment by no more than $0.25$\,G, even for the broadest resonances. 
The calculated values of $\abg \Delta$ are in good agreement with the experiment within at most two standard deviations. 

\section{Testing Efimov universalities}

Equipped with a high-precision map $a(B)$, we now turn to tests of Efimov universalities. The three-body loss rate of a thermal Bose gas is given by
\begin{equation}
\dot N/N=-L_3 n^2\,,
\label{eqNdot}
\end{equation}
which upon integration yields $1/N^2 = 2t L_3 /V^2 + 1/N_0^2$ in a uniform system of volume $V$.
The scaling $L_3\propto \hbar a^4/m$, expected on dimensional grounds (for $\avdw \ll a \ll \lambda_T$, where $\lambda_T$ is the thermal wavelength) is further modulated by Efimov physics; zero-range zero-temperature theory predicts log-periodic deviations~\cite{Braaten:2007}, captured by the dimensionless 
\begin{equation}
\vspace{-0.4em}
\mathcal{Z}_3\equiv\frac{m L_3} {3\hbar a^4}\,,
\end{equation}
\vspace{-0.4em}
where
\begin{equation}
\mathcal{Z}_3=\begin{dcases*}
      4590 \frac{\sinh(2\eta^*)}{\sin^2\big{[}s_0\ln\big{(}\tfrac{a}{a_-}\big{)}\big{]}+\sinh^2(\eta^*)}  & for $a<0$\\
      67.12 e^{-2\eta^*}\big{(}\sin^2\big{[}s_0\ln\big{(}\tfrac{a}{a_+}\big{)}\big{]}+\sinh^2[\eta^*]\big{)}\\+16.84\big{(}1-e^{-4\eta^*}\big{)} & for $a>0$\,,
    \end{dcases*}
    \label{eqZ3}
\end{equation}
with $s_0=1.00624$. For $a<0$, loss resonances occur at $a_-e^{j\pi/s_0}$, where $j\in \{0,1,...\}$,
whereas for $a>0$ the log-periodic sinusoidal modulation exhibits minima at $a_+e^{j\pi/s_0}$ and corresponding maxima at $a_pe^{j\pi/s_0}$, with $a_p=a_+e^{\pi/(2s_0)}$, such that $a_-/a_p=-1$. Here $\eta^*$ denotes the Efimov width parameter, which encodes the lifetime of the Efimov trimers.

\begin{table*}[t!]
\caption{\label{tab3}
Extracted Efimov loss features across the four intermediate-strength Feshbach resonances from \hyperref[fig1]{Fig.\,1(d)}, as well as corresponding Efimov features near ~$\Bres=33.6$\,G from Refs.~\cite{Chapurin:2019,Xie:2020}.
Note that Efimov--van-der-Waals universality predicts $a_-=-9.7\avdw\approx -630a_0$ for $^{39}$K, while the universal Efimov ratio $a_p/a_-=-1$. 
For reference, we include the approximate values of $s_{\rm res}$ from Ref.~\cite{Roy:2013} and values of the effective range $r_{\rm eff}$ (calculated close to $\Bres$) from our coupled-channel calculations.
}
\begin{ruledtabular}
\begin{tabular}{ccccccccc}
$\ket{F,m_{F}}$ & $B_{\rm res}$ (G) & $s_{\rm res}$ & $r_{\rm eff} (a_0)$& $a_- (a_0)$ & $a_-/\avdw$& $\eta^*_-$ &$a_p (a_0)$ &$a_p/a_-$\\
\hline
$\ket{1,1}$& $402.74(1)$ & $2.8$ & $136$ & $-830(40)$ & $-12.8(6)$ & $0.27(6)$ & $900(60)$ & $-1.08(9)$\\
$\ket{1,0}$& $472.33(1)$ & $2.8$ & $137$ &  $-840(30)$ & $-13.0(5)$ & $0.26(5)$ & $760(70)$ & $-0.90(9)$\\
$\ket{1,-1}$& $33.5820(14)$ \cite{Chapurin:2019} & $2.6$& $135$ & $-908(11)$~\cite{Chapurin:2019} & $-14.05(17)$~\cite{Chapurin:2019} &$0.25(1)$~\cite{Chapurin:2019} & $876(28)$~\cite{Xie:2020} & $-0.96(3)$ \cite{Xie:2020} \\
$\ket{1,-1}$& $162.36(2)$ & $1.1$ & $59$ & $-780(70)$& $-12.1(11)$& $0.5(1)$ & - & - \\ 
$\ket{1,-1}$& $561.14(1)$ & $2.5$ & $132$ & $-810(30)$ & $-12.5(5)$& $0.33(6)$ & $800(100)$ & $-0.99(13)$\\ 
\end{tabular}
\end{ruledtabular}
\end{table*}

To study the loss dynamics we prepare box-trapped thermal gases with initial densities $n_0$ ranging from $0.08\,\upmu$m$^{-3}$ to  $3\,\upmu$m$^{-3}$. We use trap depths $\UD/\kB \sim
1\,\upmu$K and restrict ourselves to temperatures $T\lesssim 100$\,nK to mitigate nonzero-$T$ effects, which are especially prominent for $a<0$ near $a_-$~\cite{Huang:2015,Wacker:2018,Chapurin:2019}.
For a given $T$ and $n_0$ we initiate the loss measurements either by slowly (in tens of ms) ramping $B$ to set a final $a$ or by performing a spin flip at the field of interest \cite{Wacker:2018}; see also \hyperref[appB]{Appendix\,B} for quench-based loss measurements featuring rich molecular dynamics and high-density effects. 

In \hyperref[fig3]{Fig.\,3(a)} we show an example of the loss dynamics for a thermal gas at $a=-930(30)a_0$ in the $|1,1\rangle$ state. Our uniform samples offer negligible anti-evaporation; heating occurs only for high $a$ when the mean free path $\ell = (8\pi n a^2)^{-1}$ becomes short, such that the loss products (with energies $\sim\Eb\gg\UD$) undergo additional collisions and deposit some of their energy before leaving the trap.
In all cases we restrict our analysis to times where $T$ remains within $\sim 10\%$ of its initial value.
The slope of the linear $1/N^2(t)$~\footnote{When extracting $L_3$, we also include effects of one-body losses (with lifetime $\sim 100\,$s) due to collisions with background-gas particles. Note that for our highest temperatures used for the low positive $a$ measurements, we observe additional slow essentially $a$-independent loss captured by a one-body lifetime $\sim 30\,$s, which we include in the same way.} yields~$L_3/V^2$, from which we calculate $\mathcal{Z}_3$~\footnote{We use a range of different-sized box traps and independently calibrate $V$ using \emph{in situ} measurements, also accounting for slight changes in $V$ with $T$ and $\UD$ owing to the finite sharpness of the trap walls.}.

In \hyperref[fig3]{Fig.\,3(b,c)} we show the extracted $\mathcal{Z}_3(a)$ on both sides of the $402.7$\,G Feshbach resonance.
For $a<0$ (b), we see an order of magnitude enhancement of $\mathcal{Z}_3$ signaling $a_-$, while for $a>0$ (c) $\mathcal{Z}_3$ is consistent with a log-periodic oscillation.
To extract the positions of the maxima of $\mathcal{Z}_3$ we fit each series with Eqs.~(\ref{eqZ3}), but also include a free amplitude $P$ to account for systematic uncertainties in density and the approximation $\langle n^2 \rangle=n^2$ made in Eq.~(\ref{eqNdot}); we find $P\approx 0.5$, consistent with 1 within our systematic uncertainties.
The solid lines display the average fit across series, while the dotted ones extend it outside of the fit range.

In \hyperref[tab3]{Table\,III} we summarize our characterization of the Efimov features for measurements across four Feshbach resonances, as well as measurements from Refs.~\cite{Chapurin:2019,Xie:2020}.
In all cases we find values of $a_-=-13(1)\avdw$, with the Efimov ratios $a_{p}/a_-\approx -1$. 
We also find little variation between the Efimov width parameters $\eta^*_-$~(extracted from resonances at $a<0$) for different states. 
Curiously, for the $|1,1\rangle$ state at $\Bres=402.74$\,G the three-body loss rate of thermal unitary gases has suggested lower values $\eta^*\approx0.1$\cite{Fletcher:2013,Eigen:2017,Eigen:2019-th}.

Table \ref{tab3} includes values of the effective range $r_\textrm{eff}$ near $B_\textrm{res}$ for each resonance, obtained from  coupled-channel calculations as described in \hyperref[appA]{Appendix\,A}. For the four stronger resonances with $s_\textrm{res}\sim2.6(2)$, $r_\textrm{eff}$ is within 
a few percent of $134\,a_0$. This is about 25\% less than the value of $\sim2.8r_\textrm{vdW}$ expected for broad resonances \cite{Gao:1998} (see also \cite{Xie:2020}). 

\section{Loss dynamics at unitarity}

Finally, we also study the loss dynamics of initially degenerate gases quenched to unitarity at different resonances.
For both a thermal and a degenerate homogeneous Bose gas at unitarity one can define a dimensionless loss rate 
\begin{equation}
\Gamma = -t_n \dot{N}/N\,,
\end{equation}
where $t_n$ is the density-set timescale $t_n = \hbar/E_n$, with $E_n = \hbar^2 (6\pi^2 n)^{2/3}/(2m)$.

For a thermal unitary gas~\cite{Rem:2013,Fletcher:2013,Eismann:2016}
\begin{equation}
    \Gamma = (1-e^{-4\eta^*}) \frac{18 \sqrt{3}}{\pi^2}\left(\frac{E_n}{E}\right)^2\,,
    \label{eq:gammath}
\end{equation}
where $E=(3/2)\kB T$, whereas for a degenerate gas $\Gamma=A$, where $A$ depends only on the Efimov physics~\cite{DIncao:2018c}. Previously, $A=0.28(3)$ was measured for $\Bres=402.7$\,G in $\ket{1,1}$~\cite{Eigen:2017} and $A\approx 0.18$ estimated from a measurement in $^{85}$Rb~\cite{Klauss:2017}, where $\eta_-^*=0.057(2)$\cite{Wild:2012}.

\begin{figure}[t!]
\centerline{\includegraphics[width=\columnwidth]{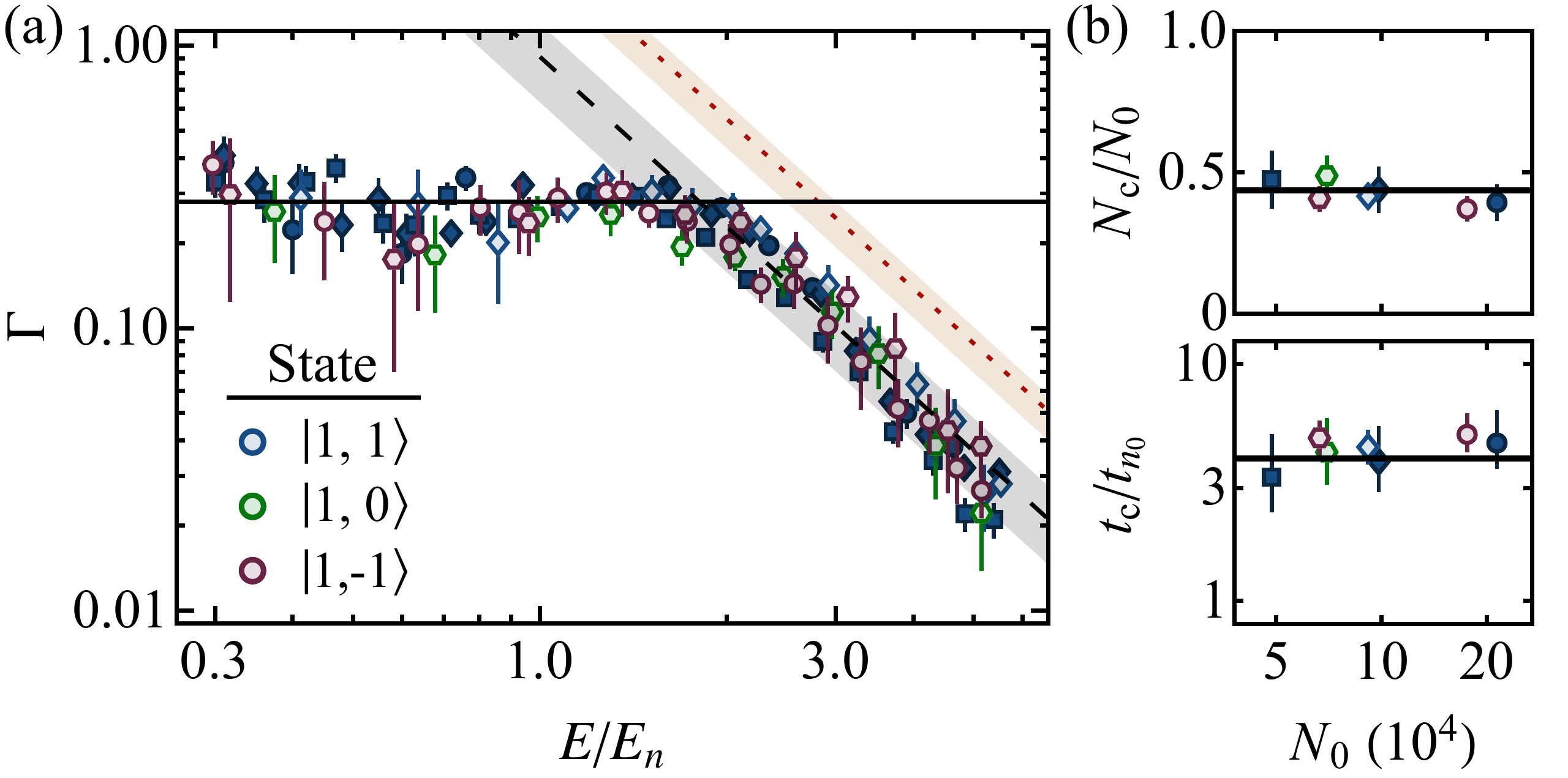}}
\caption{Comparison of loss dynamics at unitarity for the three broadest intrastate resonances, each in a different internal state. The three states are indicated by the symbol color, while different symbol shapes indicate different initial atom numbers $N_0$ [see (b)]. The three filled symbols correspond to $|1,1\rangle$ data reproduced from Ref.~\cite{Eigen:2017}. All measurements were taken using the same box trap, with $V\approx 3.5\times 10^{-14}\,$m$^3$. 
(a)~Dimensionless loss rate $\Gamma=-t_n\dot{N}/N $ versus $E/E_n$. The solid horizontal line corresponds to $\Gamma=0.28$. The dashed line corresponds to $\eta^*=0.09(4)$\cite{Fletcher:2013} in Eq.~(\ref{eq:gammath}) and the dotted line to $\eta^*=0.3(1)$.
(b) Remaining fraction of atoms at the degenerate- to thermal-gas crossover, $N_{\rm c}/N_0$ (top) and the crossover time $t_{\rm c}/t_{n_0}$ (bottom).
}
\label{fig4}
\end{figure}

As in Ref.~\cite{Eigen:2017}, we start with a quasi-pure Bose--Einstein condensate and quench it to unitarity. As the cloud decays and heats, from the time evolution of $N$ and $E$ we map out $\Gamma$ as a function of $E/E_n$, and observe a crossover from degenerate- to thermal-gas behavior.

In \hyperref[fig4]{Fig.\,4} we show $\Gamma(E/E_n)$ for the three broadest intrastate resonances [at $402.74$\,G, $472.33$\,G, and $561.14$\,G].
Consistent with observing similar values of $\eta^*_-$ from the Efimov loss resonances, we find remarkably universal behavior of $\Gamma$ across both degenerate- and thermal-gas regimes. In all cases degeneracy persists for $t_{\rm c}\approx 4t_{n_0}$, with a remaining fraction $N_{\rm c}/N_0\approx 0.4$ [see \hyperref[fig4]{Fig.\,4(b)}].

\section{Conclusions \& Outlook}
\vspace{-0.5em}
In conclusion, our experiments provide a high-precision characterization of few-body physics in $^{39}$K, offering invaluable input for modeling its full interaction landscape and understanding complex multi-channel effects~\cite{Chapurin:2019,Tiemann:2020,Secker:2021}.
Our precise determination of $a(B)$ across multiple Feshbach resonances and spin states allows a comprehensive study of Efimov physics in $^{39}$K. 
For the accessible Feshbach resonances, which all feature a similar intermediate-strength character (captured by $s_{\rm res}$), we find $a_-= -13(1)\avdw$, suggesting a universal breakdown of Efimov--van-der-Waals universality set by $s_{\rm res}$. Moreover, the observed Efimov ratios uphold Efimov universality and we find that the loss dynamics, both at and away from unitarity, are remarkably universal.

Our work points to many avenues for further research.
It would be interesting to explore the effects of few-body physics on the many-body state at unitarity\cite{Eigen:2018}. 
This includes surveying Feshbach resonances across atomic species to find cases where the unitary lifetime is longer, and understanding the intricate few-body loss channels~(see \emph{e.g.} \cite{Musolino:2022}) to devise schemes that could extend the lifetime~(as \emph{e.g.}~in molecular samples~\cite{Anderegg:2021,Schindewolf:2022}).
Finally, the rich $a(B)$ landscape of $^{39}$K also offers exciting prospects for novel experiments.
The pair of overlapping Feshbach resonances at $472.3$\,G and $491.2$\,G gives access to a steep zero crossing for studies of the collapse of uniform condensates via modulational instability, while the two the interstate resonances at $445.4$\,G and $526.2$\,G are particularly promising for creating uniform Bose mixtures.

\section{Acknowledgments}
\vspace{-0.5em}
We thank Leticia Tarruell, Eric A. Cornell, Paul S. Julienne, Nirav Mehta, and Jos\'{e} D'Incao for discussions, and Robert P. Smith for comments on the manuscript.
This work was supported by EPSRC [Grants No.~EP/N011759/1, No.~EP/P009565/1, and No.~EP/P01058X/1], ERC (UniFlat), and STFC [Grant No.~ST/T006056/1].
A.~C. acknowledges support from the NSF Graduate Research Fellowship Program (Grant No. DGE2040434). C.~E. acknowledges support from Jesus College (Cambridge). Z.~H. acknowledges support from the Royal Society Wolfson Fellowship. 

\begin{table*}[t!]
\caption{\label{tab4}
Comparison of experimental resonance parameters with those from coupled-channel scattering calculations.
`-' indicates cases where the parameter has not been determined, while `/' indicates those that do not occur.
}
\begin{ruledtabular}
\begin{tabular}{cccccccccccc}
States&\multicolumn{3}{c}{Experiment}&&\multicolumn{6}{c}{Theory}\\
$\ket{F,m_{F}}_1+ \ket{F,m_{F}}_2$ & $B_{\rm res}$ & $a_{\rm bg} \Delta$  & $B_{\rm zero}$ && $B_{\rm res}$  & $a_{\rm bg} \Delta$  & $\Delta$ & $a_{\rm bg}$ & $\Gamma_{\rm inel}$ & $B_{\rm zero}$\\
& (G)& ($a_0$\,G) & (G) & & (G) & ($a_0$\,G) & (G) & ($a_0$) & ($\upmu$G) & (G)
\\
\hline
$\ket{1,1}+\ket{1,1}$&$25.91(6)$ & - & - &  & $25.879$ & $15.39$ & $-0.465$ & $-33.10$ & 0 & -\\
$\ket{1,1}+\ket{1,1}$&$402.74(1)$ & $1530(20)$ & $350.4(1)$\cite{Fattori:2008b} & & $402.554$ & $1514$ & $-51.291$ & $-29.52$ & 0 & $350.714$\\
$\ket{1,1}+\ket{1,1}$&$752.3(1)$\cite{DErrico:2007} & - & - & & $752.255$ & $14.01$ & $-0.397$ & $-35.30$ & 0 & -\\
$\ket{1,0}+\ket{1,0}$ & $58.97(12)$  & - & - & & $58.949$ & $194.2$ & $-6.648$ & $-29.22$ & $-5$ & -\\
$\ket{1,0}+\ket{1,0}$ & $65.57(23)$ & - & - & & $65.573$ & $140.7$ & $-3.361$ & $-41.87$ & $-183$ & - \\
$\ket{1,0}+\ket{1,0}$ & $472.33(1)$  & $2040(20)$ & $393.2(2)$ & & $472.118$  & $1997$ & $-117.78$ & $-16.96$ & $-8$ & $393.635$\\
$\ket{1,0}+\ket{1,0}$ &$491.17(7)$ & $140(30)$ & $490.1(2)$ & & $490.930$  & $139.5$ & $-1.045$  & $-133.43$ & $-2$ & $489.931$\\
$\ket{1,-1}+\ket{1,-1}$& $33.5820(14)$\cite{Chapurin:2019}   & $-1073$\cite{Chapurin:2019} & / & & $33.568$ & $-1073$ & $79.469$ & $-13.50$ &  $100$ &  /\\  
$\ket{1,-1}+\ket{1,-1}$&$162.36(2)$ & $760(20)$ & / &  & $162.347$  & $711.3$ &  $-60.628$ & $-11.73$ & $-180$ &  /\\
$\ket{1,-1}+\ket{1,-1}$& $561.14(2)$ & $1660(20)$ & $504.9(2)$ & & $560.935$ & $1611$ & $-55.358$ & $-29.10$ &  $-9$ &  $504.717$\\

$\ket{1,1}+\ket{1,0}$ & $25.81(6)$ & - & - & & $25.785$ & $47.53$ & $-1.345$ & $-35.35$ & $-5$ & -\\
$\ket{1,1}+\ket{1,0}$ & $39.81(6)$ & - & - & & $39.835$ & $84.47$ & $-2.061$ & $-40.99$ & $-56$ & -\\
$\ket{1,1}+\ket{1,0}$ & $445.42(3)$ & $1110(40)$ & - && $445.293$  & $1140$ & $-37.569$ & $-30.35$  & $-9$ & -\\
\hphantom{$-$}$\ket{1,1}+\ket{1,-1}$ & $77.6(4)$ & - & - & & $77.725$ & $2618$ & $-95.738$ & $-27.35$ & $-8738$ & -\\
\hphantom{$-$}$\ket{1,1}+\ket{1,-1}$ & $501.6(3)$ & - & - & & $501.480$ & $897.8$ & $-18.304$ & $-49.05$ & $-3364$ & - \\
\hphantom{$-$}$\ket{1,0}+\ket{1,-1}$ & $113.76(1)$~\cite{Tanzi:2018} & $715(7)$~\cite{Tanzi:2018}  & - && $113.768$ & $755.7$ & $-19.215$ & $-39.33$ & $-295$  & - \\
\hphantom{$-$}$\ket{1,0}+\ket{1,-1}$ & $526.21(5)$ & $970(50)$  & - && $525.995$ & $876.8$ & $-28.249$ & $-31.04$ & $-9$ & - \\
\end{tabular}
\end{ruledtabular}
\end{table*}

\section*{Appendix A: Comparison of Feshbach resonance properties with coupled-channel calculations}
\label{appA}

\vspace{-0.5em}

We have carried out coupled-channel scattering calculations for comparison with our measurements of the Feshbach resonance properties.
The total wave function is expanded in a complete basis set of functions for electron and nuclear spins and end-over-end rotation, producing a set of coupled differential equations that are solved by propagation with respect to the internuclear distance $R$. The methods used are similar to those in Ref.\ \cite{Berninger:Cs2:2013}, so only a brief outline is given here. The Hamiltonian for the interacting pair is
\begin{equation}
\label{full_H}
\hat{H} =\frac{\hbar^2}{2\mu}\left[-\frac{1}{R}\frac{d^2}{dR^2}R
+\frac{\hat{L}^2}{R^2}\right]+\hat{H}_\textrm{A}+\hat{H}_\textrm{B}+\hat{V}(R),
\end{equation}
where $\mu$ is the reduced mass and $\hat{L}$ is the two-atom rotational angular momentum operator. The single-atom Hamiltonians $\hat{H}_i$ contain the hyperfine couplings and the Zeeman interaction with the magnetic field. The interaction operator $\hat{V}(R)$ contains the two isotropic Born--Oppenheimer potentials, for the X $^1\Sigma_g^+$ singlet and a $^3\Sigma_u^+$ triplet states, and anisotropic spin-dependent couplings that arise from magnetic dipole-dipole and second-order spin-orbit coupling. Here we use the singlet and triplet interaction potentials of Tiemann et al.\,\cite{Tiemann:2020} (their model 3), with the second-order spin-orbit coupling function of Xie et al.~\cite{Xie:2020}.

The scattering calculations are carried out using the \textsc{molscat} package \cite{molscat:2019, mbf-github:2022}.
The wavefunction is expanded in a fully uncoupled basis set that contains all allowed electron and nuclear spin functions, and is limited by $L_{\rm max}=2$. Solutions are propagated from $R_\textrm{min}=5 a_0$ to $R_\textrm{mid}=30 a_0$
using the fixed-step symplectic log-derivative propagator of Manolopoulos and Gray~\cite{Manolopoulos:1995} with an interval size of $0.002 a_0$, and from $R_\textrm{mid}$ to $R_\textrm{max}=8,000 a_0$ using the variable-step Airy propagator of Alexander and Manolopoulos~\cite{Alexander:1987}.

Each scattering calculation produces the scattering matrix $\boldsymbol{S}$ for a single value of the collision energy $E_\textrm{coll}$ and magnetic field $B$.
The complex energy-dependent s-wave scattering length $a(k_0)$ is obtained from the diagonal element of $\boldsymbol{S}$ in the incoming channel, $S_{00}$, using the identity~\cite{Hutson:2007}
\begin{equation}
a(k_0) = \frac{1}{{\rm i}k_0} \left(\frac{1-S_{00}(k_0)}{1+S_{00}(k_0)}\right),
\end{equation}
where $k_0$ is the incoming wavenumber, related to the collision energy by $E_\textrm{coll}=\hbar^2k_0^2/(2\mu)$.
 Note that $a(k_0)$ becomes constant at sufficiently low $E_\textrm{coll}$, with limiting value $a$. In the present work, s-wave scattering lengths are calculated at $E_\textrm{coll}/k_\textrm{B} = 100$~pK, which is low enough to neglect the dependence on $k_0$.

When both atoms are in their lowest state $|1,1\rangle$, only elastic scattering is possible. The scattering length is then real and approximately follows Eq.~(2) in the vicinity of an isolated resonance. If either atom is excited, however, inelastic collisions may also occur. For most of the atomic states considered here, these are spin-relaxation collisions, mediated by the weak spin-spin and second-order spin-orbit couplings. The scattering length is then complex, $a(B)=\alpha(B)-\textrm{i}\beta(B)$, and the rate coefficient for two-body loss is approximately proportional to $\beta$ \cite{Hutson:2007}. Even if there is very little inelastic scattering away from resonance, $\beta(B)$ shows a narrow peak of height $a_\textrm{res}$ near $\Bres$ and $\alpha(B)$ shows an oscillation of magnitude $\pm a_\textrm{res}/2$ instead of a pole. A feature of \textsc{molscat} is that it can converge automatically on both elastic and inelastic Feshbach resonances and characterize them to obtain their parameters, as described by Frye and Hutson~\cite{Frye:2017}. For elastic scattering, the parameters extracted are $\Bres$, $\Delta$, and $\abg$; in the presence of weak inelastic scattering, these are supplemented by $a_\textrm{res}$, which can be recast as $\Gamma_{\rm inel}=-2\abg\Delta/a_{\rm res}$ to assess the size of the region around $\Bres$ where $a(B)$ is not pole-like~\cite{Hutson:2007,Frye:2017}. 

We have characterized theoretically all the resonances for which experimental properties are given in~\hyperref[tab1]{Table\,I} and \hyperref[tab2]{Table\,II}. The calculated properties are compared with experiment in \hyperref[tab4]{Table\,IV}. The agreement is generally good. The calculated resonance positions differ from experiment by up to 0.25~G, but that is comparable to the deviations between experiment and theory in Ref.\ \cite{Tiemann:2020}. Note that the position measured here for the resonance near $403$\,G is far closer to theory than the experimental value used in Ref.\ \cite{Tiemann:2020}. The calculated values of $\abg\Delta$ are generally within one or two standard deviations of the measured ones, but the calculation breaks them down into their components $\abg$ and $\Delta$, which the present experiments cannot do without assuming that Eq.~(\ref{eqaB}) holds all the way to $B_{\rm zero}$.
For the resonances at higher thresholds, the small values of $\Gamma_\textrm{inel}$ confirm that the scattering length is pole-like until very close to $\Bres$.

\textsc{molscat} can also converge numerically upon fields $B_\textrm{zero}$ where $\alpha(B)$ is zero. These may differ from the value $B_\textrm{res}+\Delta$ implied by the approximate Eq.~(\ref{eqaB}) if $a_\textrm{bg}$ varies significantly across the width of the resonance. These values are included in \hyperref[tab4]{Table\,IV}, and are close to the experimental values even in cases where $B_{\rm zero}$ differs significantly from $B_\textrm{res}+\Delta$.  

We have also calculated values of the effective range near $B_\textrm{res}$ from the expansion
\begin{equation}
\frac{1}{a(k_0)} = \frac{1}{a} - \frac{1}{2} r_\textrm{eff} k_0^2 + \ldots
\end{equation}
using the methods of Ref.\ \cite{Blackley:2014}. The results are included in \hyperref[tab3]{Table\,III}.

\section*{Appendix B: Peculiarities of quench-based Loss Spectroscopy}
\label{appB}
\begin{figure}[t!]
\centerline{\includegraphics[width=\columnwidth]{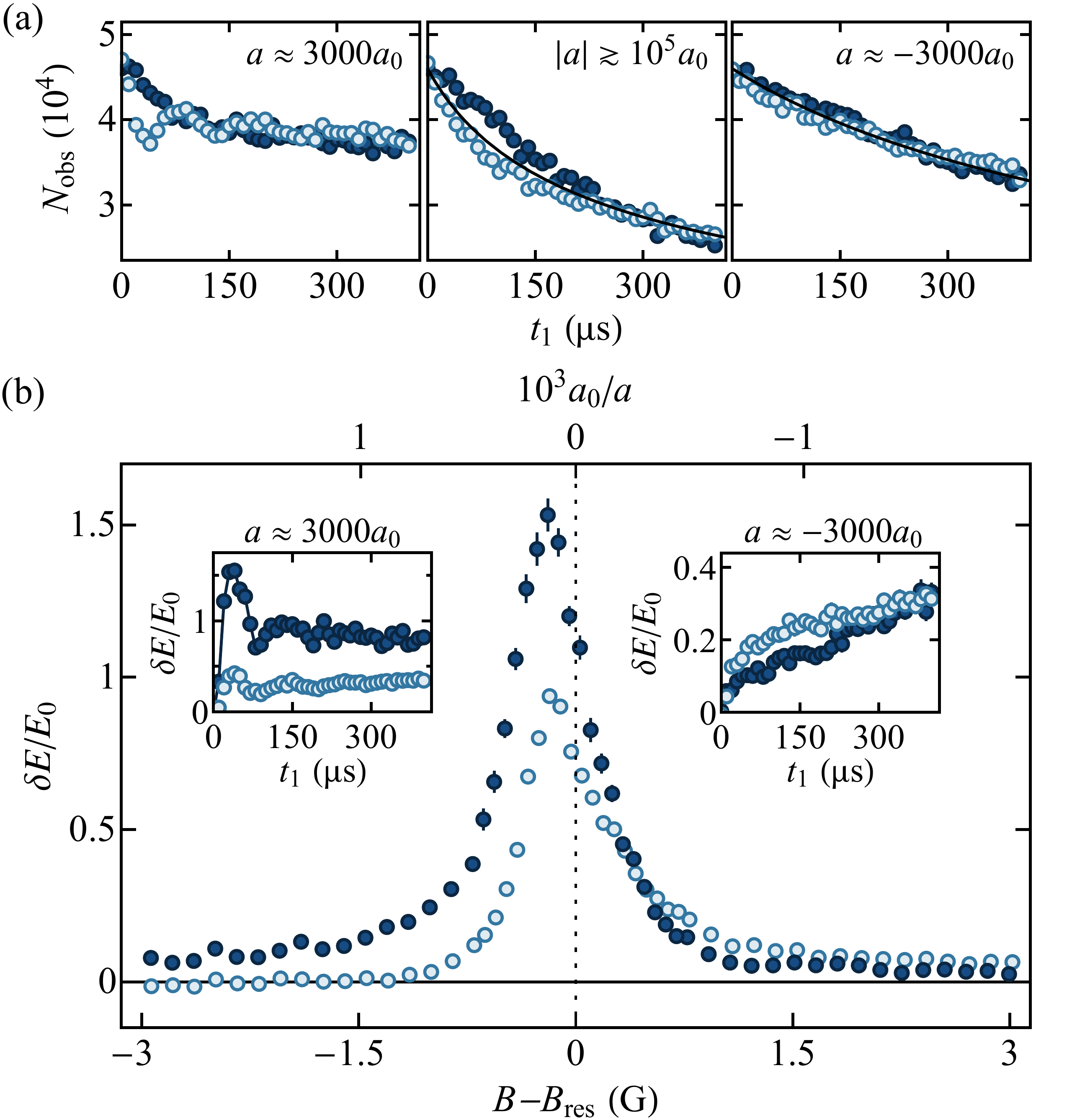}}
\caption{On the perils of quench-based loss spectroscopy [see \hyperref[fig2]{Fig.\,2(a)} for protocol]. (a)~Evolution of $N_{\rm obs}$($t_1$) for quenches below, on, and above resonance (left to right), with initial $N_0=4.6(1) \times 10^4$ atoms at $T\approx 100$\,nK. While the early times display rich dynamics, the long-time loss is independent of the quench direction.
(b)~Fractional change in energy $\delta E/E_0$ for quench-based loss spectroscopy, measured using $t_1=400\,\upmu$s [as in
\hyperref[fig2]{Fig.\,2(b)}]. Here substantial differences persist between the quench directions even at long times. We note that $\delta E/E_0$ is  not centered at $\Bres$, making it unsuitable for accurate $\Bres$ measurements.}
\label{figS1}
\end{figure}

Using field quenches for loss spectroscopy circumvents substantial atom loss in the time it takes to reach the $B$-field of interest. This leads to more narrow spectra and avoids systematic asymmetries due to the quench direction, allowing a reliable extraction of $\Bres$. However, such quenches can also induce additional dynamics, primarily dictated by shallow molecular bound states, which we explore below.

In \hyperref[figS1]{Fig.\,5(a)} we plot $N_{\rm obs}(t_1)$ for the quench experiments from \hyperref[fig2]{Fig.\,2} at three different fields, below ($B<\Bres$), on ($B\approx \Bres$), and above ($B>\Bres$) resonance. The early times reveal rich dynamics, while for late times ($t_1=400\,\upmu$s) we observe loss essentially independent of the quench direction. 

For $a\approx 3000a_0$, the first quench to $B$ already creates a non-negligible superposition of atoms and molecules, leading to an oscillation in $N_{\rm obs}(t_1)$ upon quenching back to $\approx 500\,a_0$. Meanwhile, for quenches back to $\approx-500\,a_0$, where the dimer state is unbound, $N_{\rm obs}(t_1)$ decreases monotonically. 
However, the observed loss rate is still an order of magnitude larger compared to our adiabatically prepared samples [\hyperref[fig3]{Fig.\,3(c)}].
Instead, for $a\approx -3000a_0$, we only see a minute difference between the quench directions, owing to the dimer state being unbound.
On resonance, we observe a surprisingly slow initial loss rate when quenching back to $\approx -500a_0$, where the dimer state is unbound. The loss is inconsistent with the late-time behavior that is captured by the thermal unitary atom-loss scaling law $\dot{N}/N\propto N^{26/9}$~\cite{Eigen:2017} (solid line, fit to $t_1>250\,\upmu$s).

In \hyperref[figS1]{Fig.\,5(b)} we show the change in energy per particle $\delta E/E_0$ for $t_1=400\,\upmu$s [analogous to $\delta N/N_0$ in \hyperref[fig2]{Fig.\,2(b)}]. 
Here dramatic differences persist even at long times. For $B<\Bres$, the injected $\delta E$ is significantly higher for quenches back to $\approx -500a_0$, while for $B>\Bres$ differences are less pronounced.
The maximal $\delta E$ occurs visibly below $\Bres$, arising from a complex interplay between loss and molecular dynamics, which renders it unsuitable for accurately extracting $\Bres$.

\begin{figure}[t!]
\centerline{\includegraphics[width=\columnwidth]{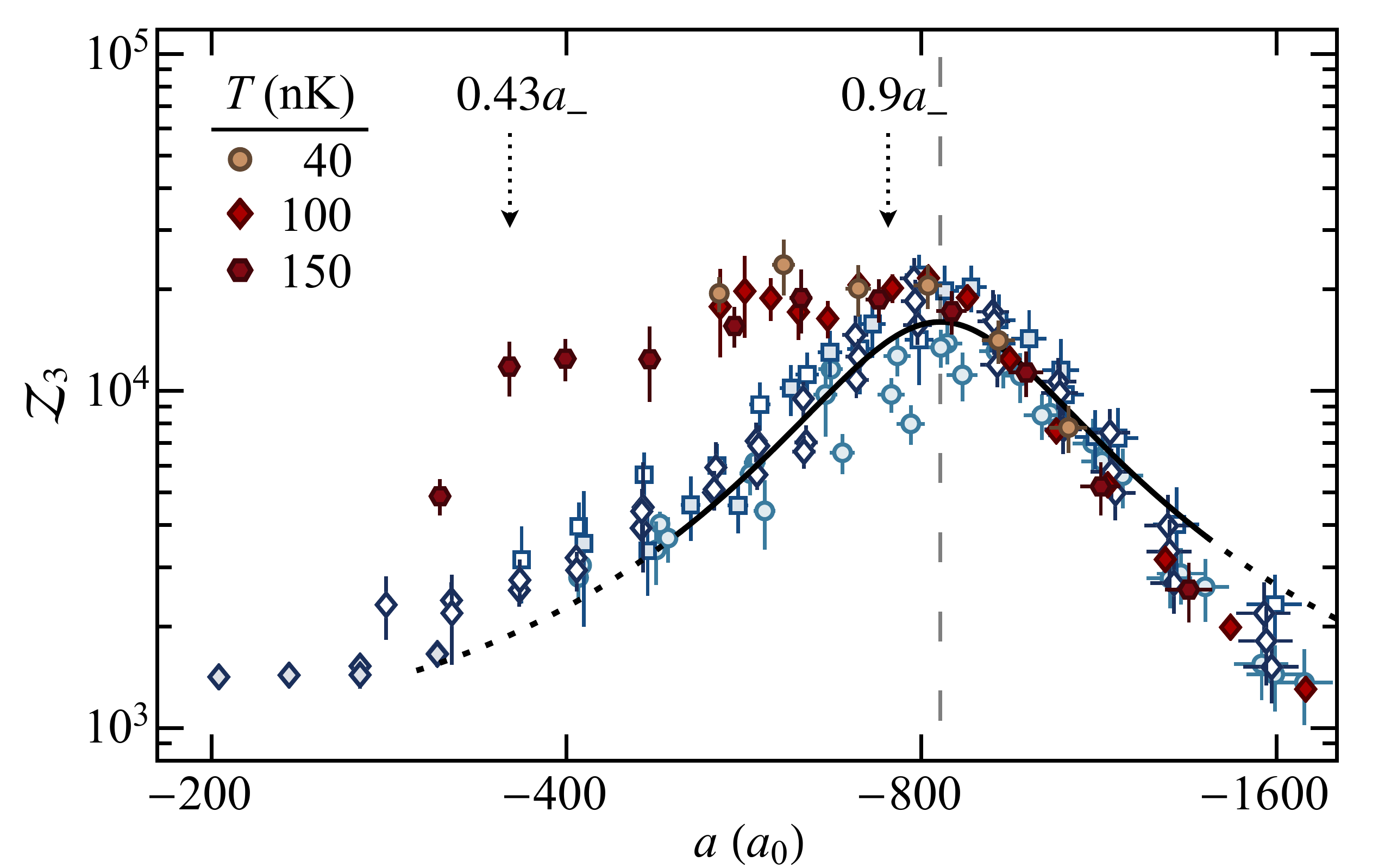}}
\caption{
Extracted $\mathcal{Z}_3$ at $a<0$ for quench-based loss spectroscopy of high phase-space density samples prepared with different $T$ (red tones, see legend) compared to data from \hyperref[fig3]{Fig.\,3(b)} (blue tones).
We posit that the significant broadening to lower $a$ occurs due to tetramer bound states related to the ground state Efimov trimer. The arrows indicate positions where these tetramer states are expected to merge with the three-atom continuum. Note that for the quench-based measurements here we scale $\mathcal{Z}_3$ between series at different $T$ to match the high-$|a|$ tail-data from \hyperref[fig3]{Fig.\,3(b)}; we attribute these $\lesssim 30\%$ differences to slight changes in $V$.
}
\label{figS2}
\end{figure}

The robust behavior of the quench-based loss dynamics for $a<0$ suggests that they might provide an accurate measurement of $\mathcal{Z}_3$.
In \hyperref[figS2]{Fig.\,6} we compare $\mathcal{Z}_3$ extracted from such quench-based measurements (red tones, shown in legend) to the data from \hyperref[fig3]{Fig.\,3(b)} (blue tones). Here we use gases with $n\lambda_T^3\sim 1$ and resolve significantly shorter timescales.
We observe significant broadening of $\mathcal{Z}_3$ towards lower $|a|$.  We posit that this might be a signature of tetramer states related to the ground-state Efimov trimer, which are predicted to merge with the three-atom continuum at $-0.43a_-$ and $-0.9a_-$~\cite{Naidon:2014}.

\section*{Appendix C: Overlapping Feshbach resonances}
\vspace{-0.0em}
\label{appC}
For two overlapping resonances with widths of the same sign, Eq.~(\ref{eqaB}) extends to the more general form~\cite{Lange:2009,Jachymski:2013}
\begin{equation}
a(B)=\abg\left( 1-\frac{\Delta_{1}}{B-B_{\rm res,1}}\right)  \left( 1-\frac{\Delta_{2}}{B-B_{\rm res,2}}\right),
\label{eqaB2}
\end{equation}
where $\abg$ is the common background scattering length, $B_{\rm res,1(2)}$ the first (second) resonance position, and  $\Delta_{1(2)}=B_{\rm zero,1(2)}-B_{\rm res,1(2)}$.
For $\Delta_1/\Delta_2\gg 1$, the effective width of the narrow resonance is $\abg' \Delta_2=\abg [1-\Delta_1/(B_{\rm res,2}-B_{\rm res,1})]\Delta_2$.

For the narrow intrastate resonance near $491\,$G we find $\abg' \Delta_2=140(30)\,a_0$G using our measurements~(see Table~\ref{tab1}), while $\abg'/\abg\approx 5.2$.
To calculate $a(B)$ for our $a_-$ measurements near the $162.4$\,G resonance we include the small ($\sim10a_0$) contribution from the $33.6$\,G one.

\begin{figure}[t!]
\centerline{\includegraphics[width=\columnwidth]{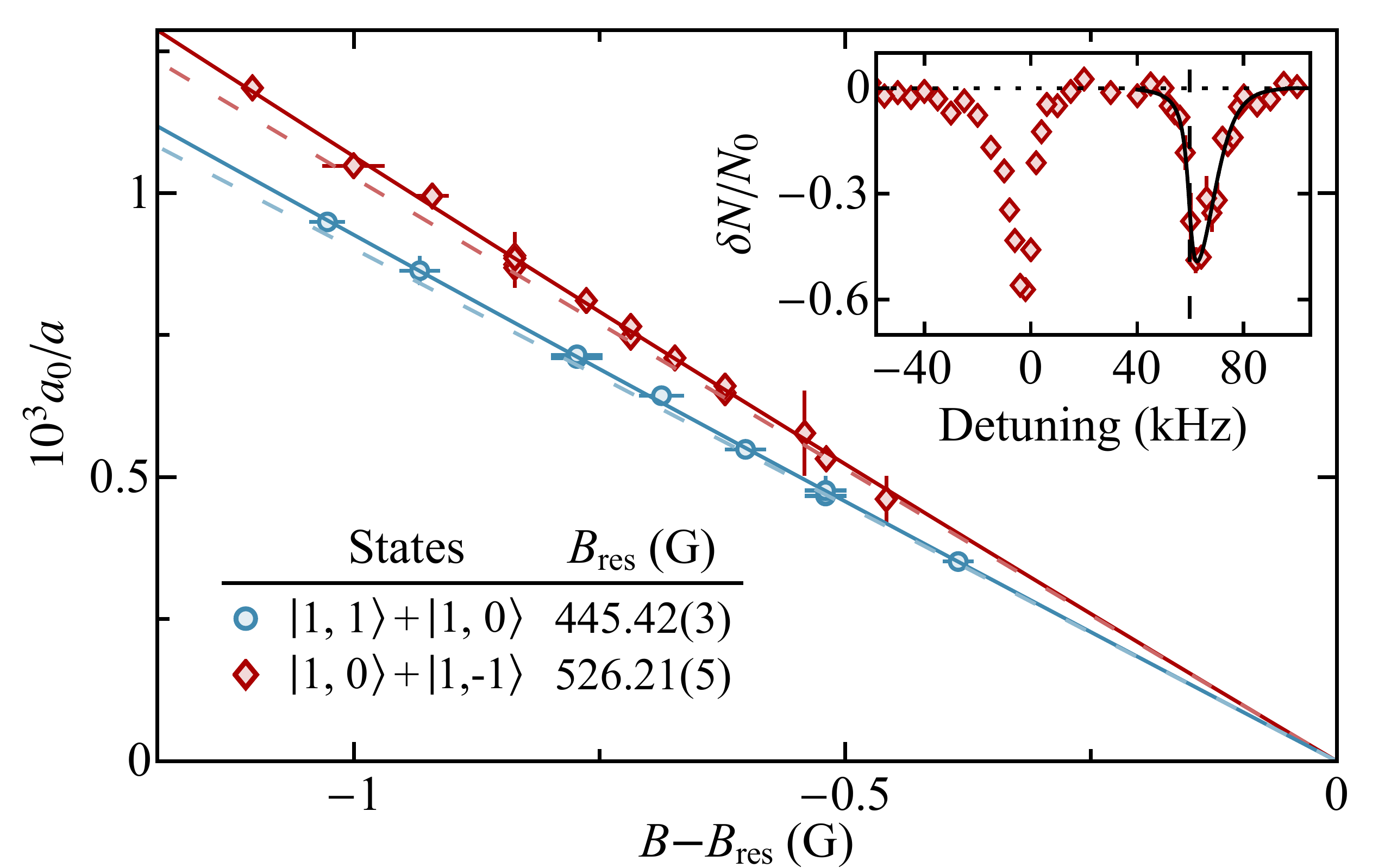}}
\caption{Inverse scattering length calculated from the extracted $\Eb$ versus $B-\Bres$. Solid lines show fits using Eqs.~(\ref{eqaB},\ref{eqEb}). The inset shows a typical rf spectrum of $\delta N/N_0$ versus rf detuning from the bare atomic transition; here the solid curve shows a fit of the dimer peak, with the extracted $\Eb/h$ indicated by the dashed line.
}
\label{figS3}
\end{figure}
\vspace{-0.5em}
\section*{Appendix D: Interstate rf association spectroscopy}
\label{appD}
\vspace{-0.5em}
Here we present $\Eb$ measurements for two interstate resonances based on rf molecular association spectroscopy~\cite{Zirbel:2008,Weber:2008,Chin:2010} [see \hyperref[figS3]{Fig.\,7}, plotted akin to \hyperref[fig1]{Fig.\,1(d)}].
To measure $\Eb$ we prepare thermal clouds spin-polarized in one of the two states, and then apply a weak rf pulse up to tens of ms to associate the interstate Feshbach dimer, which enhances atom loss.

A typical rf spectrum is shown in the inset of \hyperref[figS3]{Fig.\,7}, which exhibits features corresponding to both the bare atomic transition and the dimer.
To extract $\Eb$ (dashed line) from the asymmetric dimer peak, we fit a Lorentzian convolved with a Maxwell--Boltzmann distribution (solid line)~\cite{Weber:2008} (see also \cite{Tanzi:2018}); this asymmetry arises due to the initial kinetic energy of the associated atoms.
Curiously, the atomic feature near zero detuning also exhibits an asymmetric tail, which we also attribute to kinetic effects.
We restrict ourselves to cases where the two peaks are well separated.


%

\newpage
\cleardoublepage

\setcounter{figure}{0} 
\setcounter{equation}{0} 

\renewcommand\theequation{S\arabic{equation}} 
\renewcommand\thefigure{S\arabic{figure}}

\end{document}